\documentclass{aa}
\usepackage{graphicx}
\usepackage{txfonts}
\begin{document} 

   \title{Candidate cosmic filament in the GJ526 field, mapped with the NIKA2 camera}



\author{J.-F. Lestrade\inst{\ref{label_of_speakers_institute}}\thanks{\email{jean-francois.lestrade@obspm.fr}}
  \and  F.-X. D\'esert\inst{\ref{IPAG}}
  \and  G. Lagache\inst{\ref{LAM}}
   \and R.~Adam \inst{\ref{LLR}}
     \and  P.~Ade \inst{\ref{Cardiff}}
     \and  H.~Ajeddig \inst{\ref{CEA}}
     \and  P.~Andr\'e \inst{\ref{CEA}}
     \and  E.~Artis \inst{\ref{LPSC}}
     \and  H.~Aussel \inst{\ref{CEA}}
     \and  A.~Beelen \inst{\ref{LAM}}
     \and  A.~Beno\^it \inst{\ref{Neel}}
     \and  S.~Berta \inst{\ref{IRAMF}}
     \and  M.~B\'ethermin \inst{\ref{LAM}}
     \and  L.~Bing \inst{\ref{LAM}}
     \and  O.~Bourrion \inst{\ref{LPSC}}
     \and  M.~Calvo \inst{\ref{Neel}}
     \and  A.~Catalano \inst{\ref{LPSC}}
     \and  A. Coulais \inst{\ref{label_of_speakers_institute}}  
     \and  M.~De~Petris \inst{\ref{Roma}}
     \and  S.~Doyle \inst{\ref{Cardiff}}
     \and  E.~F.~C.~Driessen \inst{\ref{IRAMF}}
     \and  A.~Gomez \inst{\ref{CAB}} 
     \and  J.~Goupy \inst{\ref{Neel}}
     \and  F.~K\'eruzor\'e \inst{\ref{Argonne}}
     \and  C.~Kramer \inst{\ref{IRAMF},\ref{IRAME}}
     \and  B.~Ladjelate \inst{\ref{IRAME}} 
     \and  S.~Leclercq \inst{\ref{IRAMF}}
     \and  J.~F.~Mac\'ias-P\'erez \inst{\ref{LPSC}}
     \and  A.~Maury \inst{\ref{CEA}}
     \and  P.~Mauskopf \inst{\ref{Cardiff},\ref{Arizona}}
     \and  F.~Mayet \inst{\ref{LPSC}}
     \and  A.~Monfardini \inst{\ref{Neel}}
     \and  M.~Mu\~noz-Echeverr\'ia \inst{\ref{LPSC}}
     \and  L.~Perotto \inst{\ref{LPSC}}
     \and  G.~Pisano \inst{\ref{Roma}}
     \and  N.~Ponthieu \inst{\ref{IPAG}}
     \and  V.~Rev\'eret \inst{\ref{CEA}}
     \and  A.~J.~Rigby \inst{\ref{Cardiff}}
     \and  A.~Ritacco \inst{\ref{ENS}, \ref{INAF}}
     \and  C.~Romero \inst{\ref{Pennsylvanie}}
     \and  H.~Roussel \inst{\ref{IAP}}
     \and  F.~Ruppin \inst{\ref{IP2I}}
     \and  K.~Schuster \inst{\ref{IRAMF}}
     \and  S.~Shu \inst{\ref{Caltech}} 
     \and  A.~Sievers \inst{\ref{IRAME}}
     \and  C.~Tucker \inst{\ref{Cardiff}}
     \and  R.~Zylka \inst{\ref{IRAMF}}}
   \institute{
    LERMA, Observatoire de Paris, PSL Research University, CNRS, Sorbonne Universit\'e, UPMC, 75014 Paris,France
    \label{label_of_speakers_institute}
    \and
    Univ. Grenoble Alpes, CNRS, IPAG, 38000 Grenoble, France                                                                                                              
   \label{IPAG}
    \and                                                                                                                                                          
    Aix Marseille Univ, CNRS, CNES, LAM (Laboratoire d'Astrophysique de Marseille), Marseille, France 
    \label{LAM}              
    \and   
     LLR (Laboratoire Leprince-Ringuet), CNRS, École Polytechnique, Institut Polytechnique de Paris, Palaiseau, France
     \label{LLR}
     \and
     School of Physics and Astronomy, Cardiff University, Queen’s Buildings, The Parade, Cardiff, CF24 3AA, UK 
     \label{Cardiff}
     \and
     AIM, CEA, CNRS, Universit\'e Paris-Saclay, Universit\'e Paris Diderot, Sorbonne Paris Cit\'e, 91191 Gif-sur-Yvette, France
     \label{CEA}
     \and
     Univ. Grenoble Alpes, CNRS, Grenoble INP, LPSC-IN2P3, 53, avenue des Martyrs, 38000 Grenoble, France
     \label{LPSC}
     \and
     Institut N\'eel, CNRS, Universit\'e Grenoble Alpes, France
     \label{Neel}
     \and
     Institut de RadioAstronomie Millim\'etrique (IRAM), Grenoble, France
     \label{IRAMF}
     \and 
     Dipartimento di Fisica, Sapienza Universit\`a di Roma, Piazzale Aldo Moro 5, I-00185 Roma, Italy
     \label{Roma}
     \and
     Centro de Astrobiolog\'ia (CSIC-INTA), Torrej\'on de Ardoz, 28850 Madrid, Spain
     \label{CAB}
     \and
     High Energy Physics Division, Argonne National Laboratory, 9700 South Cass Avenue, Lemont, IL 60439, USA
     \label{Argonne}
     \and  
     Instituto de Radioastronom\'ia Milim\'etrica (IRAM), Granada, Spain
     \label{IRAME}
     \and
     Laboratoire de Physique de l’\'Ecole Normale Sup\'erieure, ENS, PSL Research University, CNRS, Sorbonne Universit\'e, Universit\'e de Paris, 75005 Paris, France 
     \label{ENS}
     \and
     INAF-Osservatorio Astronomico di Cagliari, Via della Scienza 5, 09047 Selargius, IT
     \label{INAF}
     \and    
     Department of Physics and Astronomy, University of Pennsylvania, 209 South 33rd Street, Philadelphia, PA, 19104, USA
     \label{Pennsylvanie}
     \and
     Institut d'Astrophysique de Paris, CNRS (UMR7095), 98 bis boulevard Arago, 75014 Paris, France
     \label{IAP}
     \and
     University of Lyon, UCB Lyon 1, CNRS/IN2P3, IP2I, 69622 Villeurbanne, France
     \label{IP2I}
     \and
     School of Earth and Space Exploration and Department of Physics, Arizona State University, Tempe, AZ 85287, USA
     \label{Arizona}
     \and
     Caltech, Pasadena, CA 91125, USA
     \label{Caltech}
   }

 \date{Received January 15, 2022; accepted January 15 , 2022}

 
  \abstract
{
Distinctive large-scale structures have been identified in the spatial distribution
of optical galaxies up to redshift $z \sim 1$.
In the more distant universe, the relationship between the dust-obscured population of star-forming galaxies observed at millimetre wavelengths 
and the network of cosmic filaments of dark matter apparent in all cosmological hydrodynamical simulations is still under study. 
Using the  NIKA2 dual-band millimetre camera, we  mapped a  
field of $\sim90$~arcminutes$^2$ in the direction of the star GJ526 simultaneously in its 1.15-mm and 2.0-mm continuum wavebands  
to investigate the nature of the quasi-alignment of five sources found ten years earlier with the MAMBO camera at 1.2~mm. We find that these sources are not clumps of a circumstellar debris disc around this star as 
initially hypothesized.  Rather, they must be dust-obscured star-forming galaxies, or 
sub-millimetre galaxies (SMGs),
in the distant background. The new NIKA2 map at 1.15~mm 
reveals a total of seven SMGs distributed in projection on the sky along a filament-like 
structure crossing the whole observed field. 
Furthermore, we show that the NIKA2 and supplemental Herschel photometric data are compatible 
with a model of the spectral energy distributions (SEDs) 
of these sources when a common redshift of 2.5 and  typical values of 
the dust parameters for SMGs are adopted. Hence, we speculate
that these SMGs might be located in a filament of the distant `cosmic web'.
The length of this
candidate cosmic filament crossing the whole map is at least 4~cMpc (comoving), 
and the separations between sources are between 0.25~cMpc
and 1.25~cMpc at this redshift, in line with expectations from cosmological simulations.
Nonetheless, further observations  to determine the precise spectroscopic redshifts of these sources are required to
definitively support this hypothesis of SMGs embedded in a cosmic filament of dark matter. 
}

\keywords{galaxies: evolution – Galaxy: formation – galaxies: groups – large-scale structure of Universe –
galaxies: clusters: intracluster medium}

\titlerunning{Candidate cosmic filament}
\maketitle

%

\section{Introduction}

Sub-millimetre Galaxies (SMGs) found in submillimetre- and millimetre-wavelength surveys 
are forming stars at prodigious rates, sometimes 
exceeding  1000~$M_{\odot}$/yr, and are enshrouded in dust with bright far-infrared luminosities $>10^{12}\,L_{\odot}$ 
(\citealp[e.g.][]{Hodg20}). Thus, SMGs  have only faint or undetectable
optical counterparts but probe the most active cosmic epoch of galaxy formation at $z \cong 2$.

Theory predicts that galaxy formation preferentially occurs along large-scale
dark matter filamentary or sheetlike overdense regions that are related to initial fluctuations
of the density field in the primordial universe \citep{Bond96} 
and formed the `cosmic web' as depicted by cosmological hydrodynamical simulations. 
It is thought that galaxy growth and high star formation 
at an early stage of cosmic evolution is critically 
tied to massive haloes of dark matter at the knots of this network of filaments \citep{Beth13}. 
A large fraction of the neutral hydrogen from the inter
galactic medium falls into these filaments  and constantly replenishes 
the gas reservoirs of galaxies through cold gas accretion;  
the cold gas streams along the filaments, survives the shocks at the virial
radii of the dark matter haloes and infalls onto the galaxies to fuel their star formation and growth 
\citep{Kere05, Deke09}. 

Definitive observational confirmation of this mechanism $-$
cold gas accretion  along filaments of the cosmic web $-$ is actively sought. 
Photoionized Lyman-$\alpha$ nebulae around quasars have been observed with 
morphologies and kinematics suggestive of intergalactic gas filaments and gas infall
\citep{Cant14, Henn15,Mart15, Mart19} but  alternative 
scenarios 
based on outflows cannot be ruled out.  
Recently, extended Ly-$\alpha$ emission plausibly powered by collisional ionization 
along filamentary structures has been observed.   
\citet{Umeh19} have discovered two Ly-$\alpha$ filaments more than 1~Mpc in extent within the protocluster SSA22 at $z=3.1$
and, remarkably, also detected embedded SMGs and X-ray sources that are distributed all along these two filaments.
\citet{Baco21} report the discovery of diffuse extended Ly-$\alpha$ emission from redshift 3.1 to 4.5, tracing 
cosmic web filaments on scales of $2.5-4$~Mpc in the MUSE Extremely Deep Field, which partially overlaps 
the Hubble Ultra Deep Field.
Also, \citet{Dadd21} report on a plausible observation of cold gas accretion streams towards the massive potential well
of the protocluster RO-1001 at $z=2.91$ and on the detection of three filaments apparently connecting
to the associated 300~kpc-wide Ly-$\alpha$ nebula. 
In cosmological simulations, the relationship between SMGs 
and the network of cosmic filaments of dark matter is debated \citep{Mill15}.

Nowadays, a region of the sky several hundreds of arcseconds in size, corresponding  to several cMpc (comoving) at redshift $z~>~1$, 
can be efficiently mapped at millimetre wavelengths using a modern camera such as NIKA2 installed on the IRAM 30-metre telescope \citep{Adam18}. 
In maps covering such a scale, the spatial distribution of SMGs  can be compared to 
the cosmic web in cosmological simulations.     
We present NIKA2 data that are a serendipitous contribution to this endeavour.

In 2007, as part of a survey designed to search for new circumstellar discs around 50 stars, 
 the field in the  direction of the nearby star GJ526 was mapped in the 1.2-mm continuum waveband using
the MAMBO camera installed on the IRAM 30-metre telescope and a quasi-alignment of 
five sources along with this star was found in projection on the sky. At the time, it was not possible  to conclude whether 
these sources were background SMGs or part of a large, clumpy circumstellar disc seen edge-on and centred on this star
 \citep{Lest09}. In 2017, we again observed the same field with the  
NIKA2 camera to test whether these sources had remained fixed and were in the background, or whether they had
moved at the same rate and in the same direction as the star, and were clumps in a circumstellar disc.
Based on the proper motion of the star, this possible displacement was 
estimated to be as large as $22.6''$ in the south-west direction after ten years.

In this paper, we report our finding with NIKA2 that the  MAMBO sources did not move  
over this ten-year period and so they must be SMGs in the distant background. 
In Sect.\,\ref{obs} we present the NIKA2 observations, and in Sect.\,\ref{map} we present the NIKA2 maps 
at 1.15~mm and 2.0~mm, and discuss the overdensity of sources in the field. In Sect.\,\ref{sed} we discuss the nature of the sources 
characterized by their spectral energy distributions (SEDs) constrained with our NIKA2 photometry and Herschel supplemental data. Finally, in 
Sect.\,\ref{diss} we discuss the remarkable spatial distribution of these sources in the 1.15-mm NIKA2 map 
along a filament-like structure and a possible association with the cosmic web of the cosmological simulations.

\section{Observations} \label{obs}

The observations were conducted with the NIKA2 dual-band millimetre
camera installed on the IRAM 30-metre telescope  at Pico Veleta (altitude 2850m) in Spain 
\citep{Monf11,Cata14,Calv16,Bour16,Adam18,Pero20}. NIKA2 operates with kinetic inductance detectors (KID)
arranged into three arrays~: two with a maximum transmission around 260~GHz (1.15~mm) with 
1140 detectors in each polarization, and one at 150~GHz (2.0~mm) with 616 detectors at a single polarization. 
The transmission bands of the detectors are broad with a $\sim50$~GHz full width at half maximum (FWHM) transmission at both wavelengths.
The combination of its large field of view (6.5 arcminutes) and high sensitivity
provides the NIKA2 camera with a high mapping speed (111
and 1388 arcminutes$^2$~mJy$^{-2}$~h$^{-1}$ at 260~GHz and 150~GHz, respectively, at zero atmospheric opacity~; \citealp{Pero20}).
The angular resolutions (main beam FWHM) are 17.7 and 11.2 arcseconds at 150 and
260 GHz, respectively.

The data presented in this study were obtained during the science verification 
observations of the NIKA2 camera on 24\ April 2017, on 14, 17, 18, 19, and 20 February 2018, and on 14\ March 2018, 
in average weather conditions (atmospheric opacity 
between 0.10 and 0.25 at 2~mm, and between 0.20 and 0.45 at 1~mm) and at elevations  
between 33$^{\circ}$ and 67$^{\circ}$.  The observing strategy
consisted of a series of $9' \times 4.5'$ on-the-fly scans with a scanning
speed of 40''/sec along two orthogonal orientations ($-60^{\circ}$, $+30^{\circ}$) in the AZ-EL 
coordinate system of the telescope to minimize residual stripping patterns in the maps.
We defined the map centre at the J2000 coordinates of the star GJ526
as in the initial MAMBO map (RA : 13h45m44.52s and DEC : $+14^{\circ}53'20.6''$). 
These observations are split in 167 scans that correspond to a total of 
9.3 hours of observations on source.
For calibration purposes, skydips were conducted to determine the coupling of each KID 
to the sky emission. An opacity correction could then be computed along the line of sight of each observation scan.

The time-ordered information (TOI) of each detector was processed in order to
remove the contribution of atmospheric and electronic correlated
noises and to obtain the maps at both wavelengths following the
procedure initially described in \citet{Adam14},  and modified in Ponthieu et
al. (in preparation). In short, an iterative process between maps and TOI was
carried out until the maps did not change anymore (less than $0.1\sigma$) after three iterations. For each
sub-scan, the first four harmonics at 2~mm (five at 1~mm), 
 each made of one sine and one cosine, were fit to the TOI with a least-square
method and removed. In more detail,
there was no masking of the source but rather a downweighting of large signal-to-noise excursions. 
This downweighting becomes more accurate when the map is close to the final map after three  iterations. 
On the first iteration, the initial map was set to zero.
The process includes some inhomogeneous weighing of the data samples 
and so leads to some large spatial scale filtering.
Thus, the procedure is designed for
optimized detection of point-sources and not extended sources. 
Nonetheless, the more aggressive filtering and the avoidance of masks in this procedure
induce a filtering effect that has to be
determined by simulations of point sources injected into the data. Thus, we established that
a photometric correction of 25~\%  that
is constant for the range of S/Ns displayed by the detected sources in our study must be applied.
It should be noted that 
this iterative part of the pipeline used to process the data is an addition
to the standard pipeline used by \citet{Pero20}. Finally, we add that
the noise of each sample was propagated to the map of each scan with a standard
inverse variance weighing. A global rescaling of the noise for each
map was done to make 
the standard deviation of S/Ns in noise map unity (see Sect.\,\ref{map}),  
in order to measure an accurate S/N for point sources.  

Point sources were detected by fitting the map with a Gaussian point spread function, in other words effectively 
applying a noise-weighted matched-filtering of the map with a fixed width Gaussian.
Position accuracy is better than 3~arcseconds (for each coordinate) for sources detected above
$4\sigma$. The photometric correction of 25~\%  described above was applied to the fit amplitude. 
The flux density scale was based on specific observations of the primary calibrator
Uranus to calibrate all individual detectors of each array and on
frequent standard observations of the secondary calibrator MWC349 to check 
the stability of this calibration over a NIKA2 run of several days.   
The resulting absolute photometric accuracy is of the order of 15~\%.
Although the weather conditions were not optimal, the expected NIKA2
sensitivity at zero opacity ($30\pm3$ and $9\pm1$ mJy$.\sqrt{s}$ at 1.15~mm and 2.0~mm, respectively~; \citealp{Pero20}) 
was recovered once the opacity was taken into account.

\begin{figure}[!h]
 \centering
   \includegraphics[width=9cm]{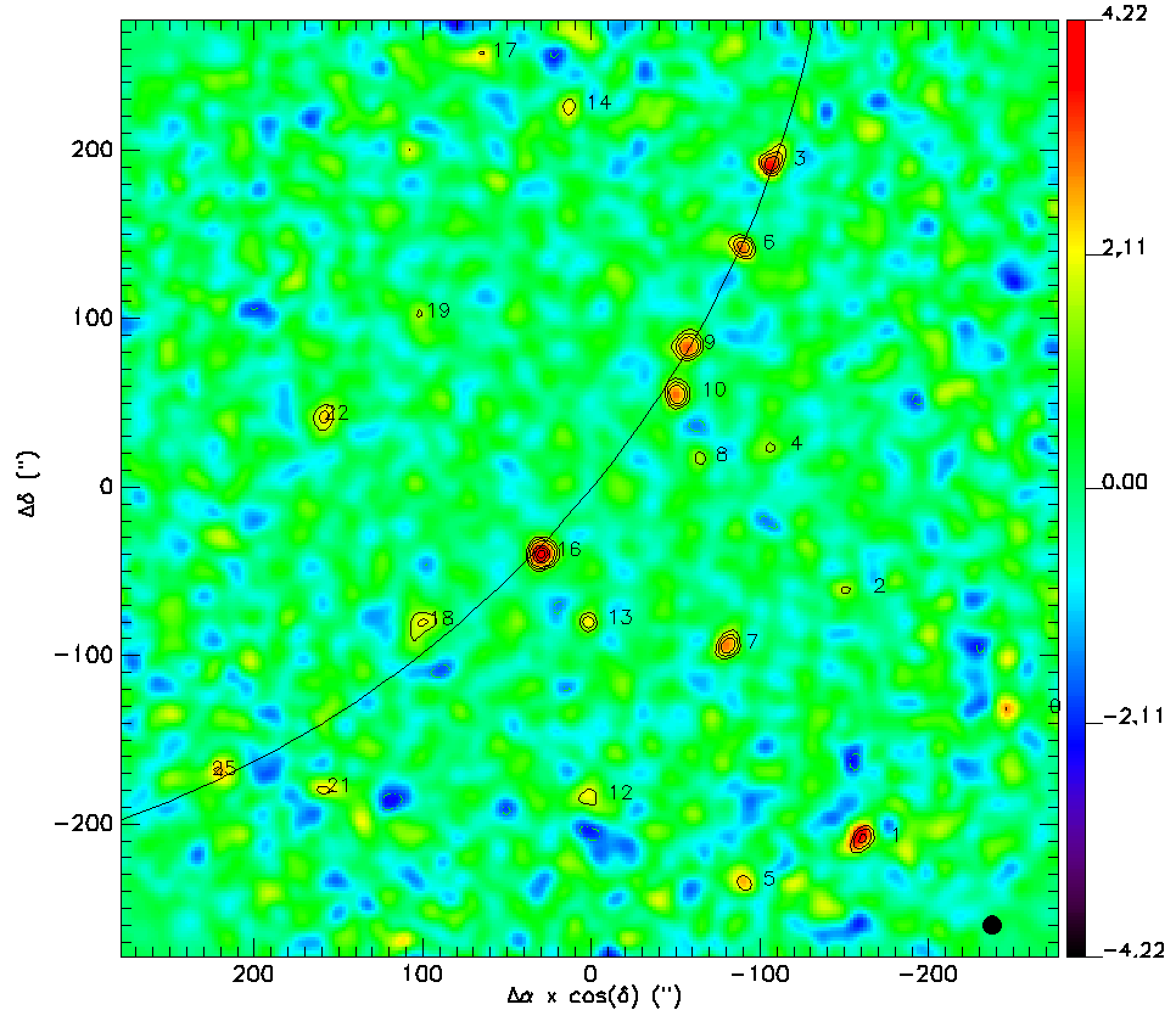}
   \caption{
GJ526 field, mapped with NIKA2 in the 1.15-mm continuum waveband (colour scale : mJy/beam, contours : S/N).
Source IDs are those in Table~\ref{tab:Tab_resul_1mm}. The curve traces the filament-like structure identified across the map. 
S/N contours are $ -3, +3, +4, +5,  +6, ....$ (the negative contours
are dashed white lines and can be seen better 
when the map is zoomed in the electronic version. The S/N=$-4$ level was processed in plotting the map but no such negative 
feature was apparent. 
For the noise rms per pixel, we refer the reader to Fig.\,\ref{fig:map_rms}. 
 Pixel size is 2 arcseconds.
J2000 coordinates of the map centre are 13h45m44.52s and $+14^{\circ}53'20.6''$. The filled black circle in the bottom right corner 
shows the HPBW.}
\vspace{1.cm}
\includegraphics[width=9cm]{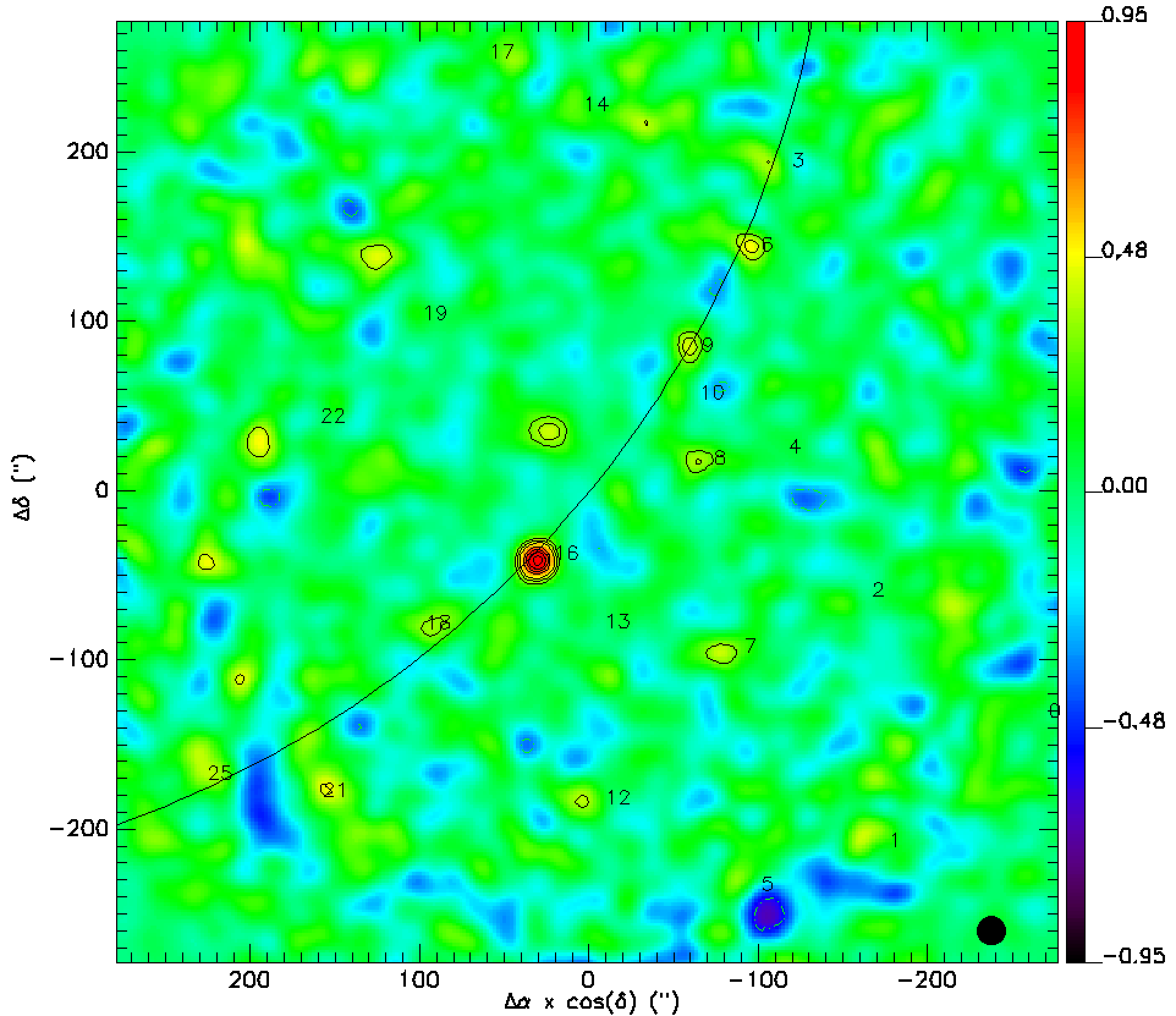}
\label{fig:map_1mm}
  \caption{Same as for Fig.\,\ref{fig:map_1mm}, but observed at 2.0~mm.}
\label{fig:map_2mm}   
\end{figure}
\begin{figure}[!h]
   \centering
   \includegraphics[width=7cm]{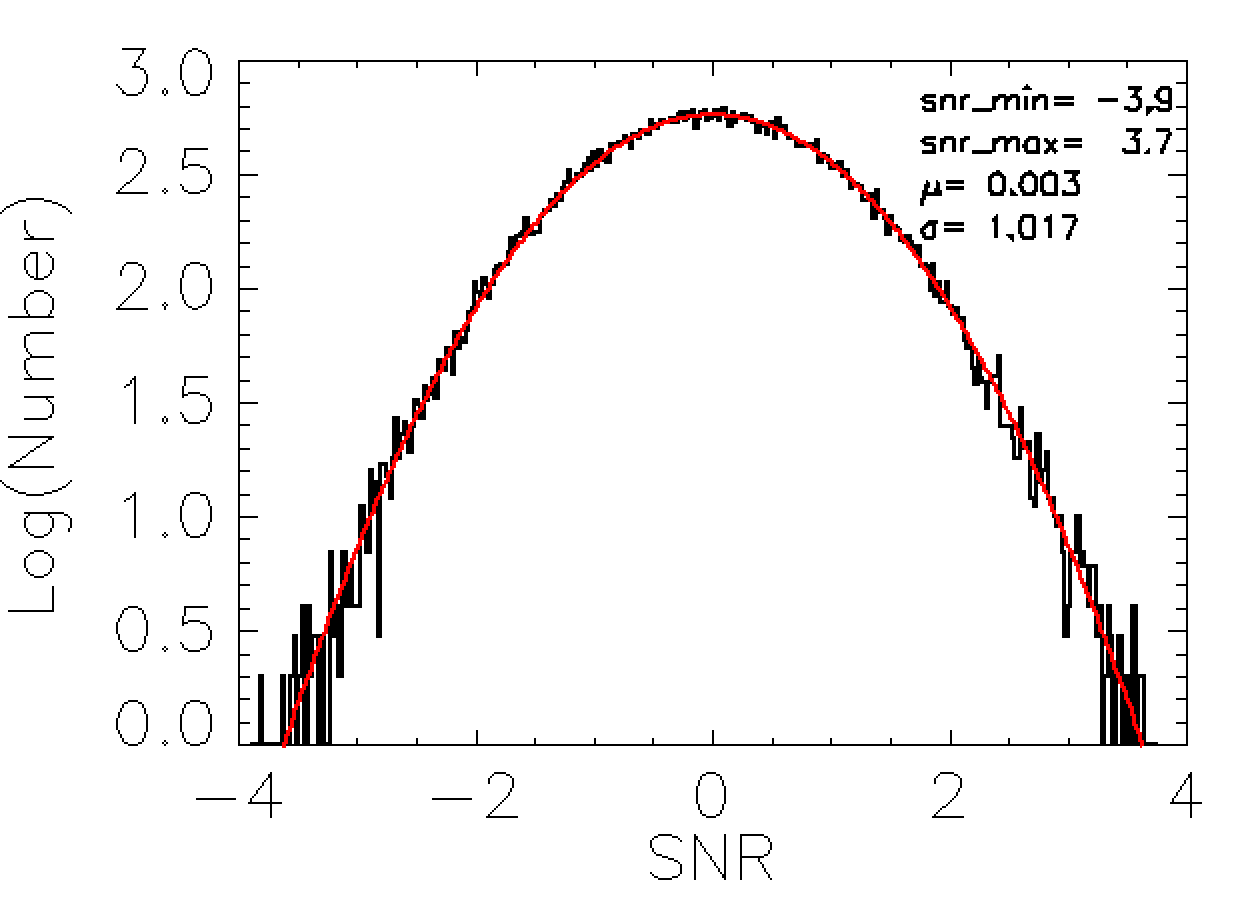}
   \caption{ S/N histogram of the jack-knifed map at 1.15~mm. The mean $\mu$ and standard deviation $\sigma$, as well as
the minimum and maximum of the S/Ns, are indicated in the figure.
In red, a Gaussian not fit but set with zero mean, standard deviation unity, and its integral summing up to the total number of S/Ns,
is overplotted to show that the theoretical distribution of S/Ns for Gaussain noise satisfactorily matches\textbf{} the histogram of the data in the map.}

          \label{fig:SNR_1mm}
\vspace{1.cm}
   \includegraphics[width=7cm]{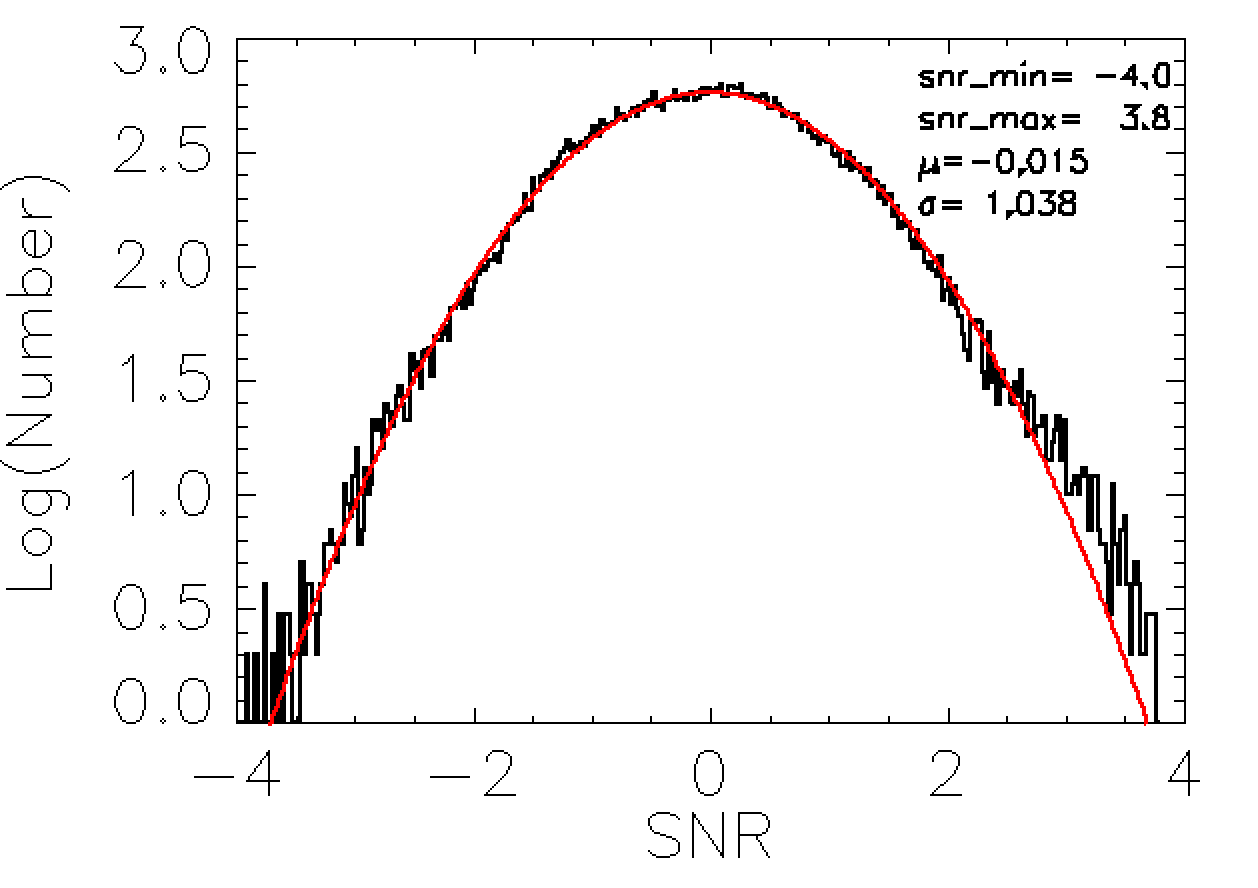}
   \caption{Same as for Fig.\,\ref{fig:SNR_1mm}, but observed at 2~mm. The slight excess
 found at the positive tail is discussed in the text.}
          \label{fig:SNR_2mm}
\vspace{1.cm}
  \includegraphics[width=7cm]{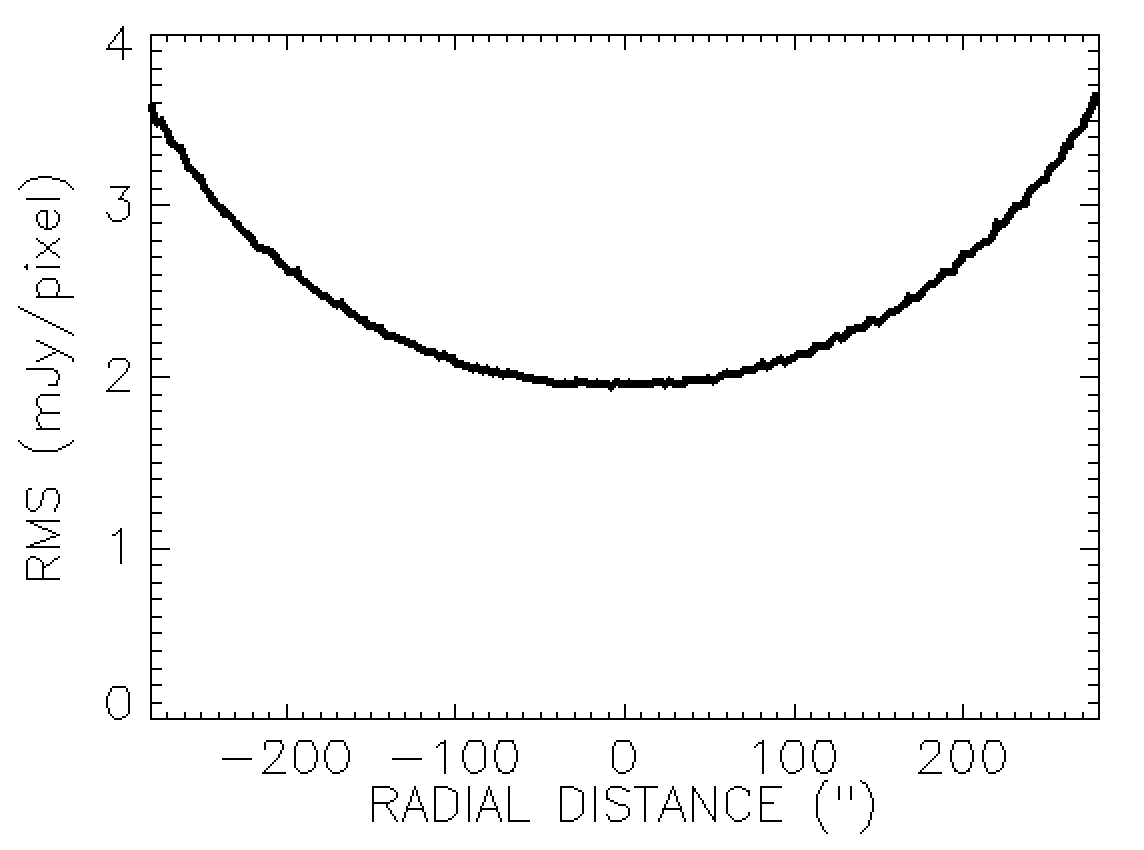}
   \caption{Radial variation in the noise rms per pixel in the 1.15-mm map (pixel size = $2''$). 
          At 2mm, the variation is the same but reduced by a factor of
roughly three. The variation is isotropic to a good approximation
          over the maps.  }
   \label{fig:map_rms}                                                                                            
\end{figure}

\section{NIKA2 maps}\label{map}

The  NIKA2 maps of the GJ526 field  $560'' \times 560''$ in size  are shown in Fig.\,\ref{fig:map_1mm} at 1.15\,mm 
and in Fig.\,\ref{fig:map_2mm} at 2.0\,mm. Millijansky point-like sources are apparent at both wavelengths.
In order to determine the detection threshold, we analysed the noise distribution in computing S/Ns at all pixels 
of the jack-knifed map at each wavelength. In such a map of the noise, the sources were eliminated after subtraction between 
individual maps of consecutive 
pairs of scans with the two orthogonal scanning orientations and finally the co-addition of all pairs.  
At 1.15~mm, we found that the S/N distribution is Gaussian, centred almost on zero 
and has a standard deviation of almost unity, as expected for normally distributed noise (see Fig.\,\ref{fig:SNR_1mm}).
However, at 2~mm, although the distribution is also centred close to zero and has a standard deviation close to unity, 
there is an excess at the positive tail that makes it deviate slightly from Gaussian, despite 
careful reprocessing (see Fig.\,\ref{fig:SNR_2mm}). This was unexpected since
the atmospheric fluctuations are less severe at this wavelength.
Close inspection of the data showed that electronic correlated noise
in the 2.0~mm sampling modules of the instrument was higher than usual
during the observations. With this caveat in mind, source extraction and photometry were carried out with caution.
        
To set the source detection threshold, 
we searched for any possible detections in the jack-knifed maps using the same fitting 
procedure of a Gaussian point spread function as described in Sect.\,\ref{obs}. Clearly, any statistically significant detection in this noise-only map 
would be spurious. At 1.15~mm, we find that there are 
nine  spurious detections with S/N~$>~+3$ and ten with S/N~$<~-3$, evenly distributed over the jack-knifed map, and none with  |S/N|~$\ge~4$.
Similarly, at 2.0~mm, there are eight spurious detections with S/N~$>~+3$ and nine with  S/N~$<~-3$, and none with  |S/N|~$\ge~4$.
We conclude that the threshold S/N~=~4 is a reliable criterion for a robust source detection
at both wavelengths. 

In Table~\ref{tab:Tab_resul_1mm} we list the 21 sources with S/N~$\ge~3$ found in the 1.15-mm NIKA2 map,
among which ten are unambiguously detected according to our criterion (S/N~$\ge~4$) and have flux densities 
between $\sim$2 and $\sim$5 mJy. 
Statistically, about two additional sources should be real among the remaining eleven sources ($3~\le~{\rm S/N}~<~4$) 
since only nine positive 
detections are expected to be spurious above $+3\sigma$, as explained above.  
Evidently, these potential sources at 1.15~mm cannot be identified unless additional inputs are included.
For instance, ID~8 is a $3.6\sigma$ source in the  1.15-mm NIKA2 map 
but can also be identified in the MAMBO map at slightly 
above $3\sigma$ \citep[see map in Fig.~1 of][]{Lest09} ; thus ID~8 is likely a real source at the level of 4.7$\sigma$ in 
quadratically combining the two S/Ns. 
Among the unambiguously detected sources at 1.15~mm, only two are also unambiguously detected 
at 2~mm (ID~16 and  ID~9), and four additional sources have S/Ns of between 3 and 4 at 2.0~mm 
(see Table~\ref{tab:Tab_resul_1mm}).
 This deficit of detections at 2~mm
can be explained by the  dust emissivity index $\beta$ of SMGs that is
between 1 and 2, or sometimes steeper (see also  Sect.\,\ref{sed}). Finally, there is a source 
which is slightly above $4\sigma$ at 2.0~mm (north-east of map and see Table~\ref{tab:Tab_resul_1mm}) 
but has surprisingly no 1.15~mm counterpart. 
As already mentioned, electronic correlated noise higher than standard during the observations 
must explain this deviation from normal statistics in the 2.0~mm map, which we consider not as reliable as the 1.15-mm map.
  
Finally, we address flux boosting that alters the measured flux densities provided in Table~\ref{tab:Tab_resul_1mm}.  
In a flux-limited survey, it is well known that sources with a low S/N are detected at flux densities 
systematically higher than their true flux densities and this effect is exacerbated when the source population increases
in number with decreasing flux \citep[e.g.][]{Copp05}. We estimated flux boosting 
with 1000 simulations of the sky in our NIKA2 maps following the prescriptions by \citet{Scot08} with their semi-Bayesian approach.
We find that it is statistically less than 15\% for a S/N above 4, and steadly increases up to 50\% for a S/N of 3, similarly
at 1~mm and 2~mm.
We chose not to correct downwards the quoted flux densities in Table~\ref{tab:Tab_resul_1mm}.


\begin{table*}[]
\center
\caption{NIKA2 astrometry and photometry. Source IDs assigned in column 1 are used throughout the text. }
\begin{tabular}{rccccccc} 
\hline \hline
         ID & Name & $\alpha_{2000}$   & $\delta_{2000}$     & $S_{260GHz}$ & $S/N$   &  $S_{150GHz}$   & $S/N$        \\ 
            &      &  $(h~~m~~s)$      &  ~$(^\circ~~'~~'')$ &   (mJy)      &  260GHz &     (mJy)       &  150GHz      \\ 
\hline \hline
\multicolumn{8}{c}{Unambiguously detected NIKA2 sources (S/N~$\ge$~4 at 1.15~mm) }   \\ 
\hline 
         16 & NIKA2$\_$J134546.6+145239  &   13:45:46.63  &     14:52:39.7     &  4.11$\pm$ 0.43 &  9.6  &  0.98$\pm$ 0.09 & 10.3 \\  
          1 & NIKA2$\_$J134533.4+144950  &   13:45:33.48  &     14:49:50.5     &  3.96$\pm$ 0.75 &  5.3  &  0.41$\pm$ 0.17 &  2.4 \\ 
          3 & NIKA2$\_$J134537.1+145631  &   13:45:37.18  &     14:56:31.9     &  3.67$\pm$ 0.60 &  6.1  &  0.39$\pm$ 0.13 &  3.0 \\ 
          9 & NIKA2$\_$J134540.6+145442  &   13:45:40.61  &     14:54:42.8     &  3.13$\pm$ 0.45 &  6.9  &  0.42$\pm$ 0.10 &  4.3 \\ 
         10 & NIKA2$\_$J134541.0+145414  &   13:45:41.09  &     14:54:14.8     &  2.97$\pm$ 0.44 &  6.8  &  0.23$\pm$ 0.09 &  2.4 \\ 
          6 & NIKA2$\_$J134538.3+145542  &  13:45:38.36   &     14:55:42.2     &  2.95$\pm$ 0.52 &  5.7  &  0.37$\pm$ 0.11 &  3.3 \\ 
          7 & NIKA2$\_$J134538.9+145145  &  13:45:38.97  &     14:51:45.4     &  2.87$\pm$ 0.47 &  6.1  &  0.41$\pm$ 0.11 &  3.9  \\ 
         22 & NIKA2$\_$J134555.5+145401  &  13:45:55.51  &     14:54: 1.1     &  2.40$\pm$ 0.52 &  4.6  &  0.02$\pm$ 0.11 &  0.2  \\ 
         18 & NIKA2$\_$J134551.5+145159  &  13:45:51.51  &     14:51:59.0     &  2.15$\pm$ 0.48 &  4.5  &  0.34$\pm$ 0.11 &  3.1  \\ 
         13 & NIKA2$\_$J134544.6+145159  &  13:45:44.69  &     14:51:59.4     &  2.08$\pm$ 0.44 &  4.7  &  0.06$\pm$ 0.10 &  0.6  \\ 
\hline \hline
\multicolumn{8}{c}{Possibly detected NIKA2 sources (3~$\le$~S/N~$<$~4 at 1.15~mm)} \\ 
\hline 
          0 &     &  13:45:27.41  &      14:51: 6.8  &  2.68$\pm$ 0.79 &  3.4 &  0.20$\pm$ 0.18 &  1.1 \\ 
         25 &     &  13:45:59.81  &      14:50:29.8  &  2.61$\pm$ 0.79 &  3.3 &  0.38$\pm$ 0.18 &  2.1 \\ 
          5 &     &  13:45:38.33  &      14:49:24.6  &  2.57$\pm$ 0.71 &  3.6 &  0.28$\pm$ 0.16 &  1.7 \\ 
         17 &     &  13:45:49.08  &      14:57:36.4  &  2.25$\pm$ 0.72 &  3.1 &  0.14$\pm$ 0.15 &  0.9 \\ 
         14 &     &  13:45:45.51  &      14:57: 5.0  &  2.20$\pm$ 0.61 &  3.6 &  0.27$\pm$ 0.13 &  2.0 \\ 
         21 &     &  13:45:55.42  &      14:50:19.7  &  2.20$\pm$ 0.67 &  3.3 &  0.48$\pm$ 0.15 &  3.1 \\ 
         12 &     &  13:45:44.66  &      14:50:15.4  &  2.15$\pm$ 0.56 &  3.8 &  0.41$\pm$ 0.13 &  3.2  \\ 
          2 &     &  13:45:34.13  &      14:52:18.1  &  1.67$\pm$ 0.52 &  3.2 &  0.16$\pm$ 0.11 &  1.4 \\ 
         19 &     &  13:45:51.63  &      14:55: 1.6  &  1.57$\pm$ 0.49 &  3.2 &  0.15$\pm$ 0.11 &  1.4 \\ 
          4 &     &  13:45:37.30  &      14:53:43.1  &  1.55$\pm$ 0.45 &  3.4 &  0.08$\pm$ 0.10 &  0.8  \\ 
          8 &     &  13:45:40.15  &      14:53:36.2  &  1.52$\pm$ 0.43 &  3.6 &  0.39$\pm$ 0.09 &  4.1  \\ 
\hline 
\end{tabular}
\label{tab:Tab_resul_1mm}
\end{table*}


\begin{table}[]
\center                                                                                                                                                                                                
\caption{Comparison between NIKA2 and MAMBO}
\begin{tabular}{ccccc}                                                                                                                                                                           
\hline \hline
\multicolumn{1}{c}{ID}  & \multicolumn{1}{c}{$NIKA2$}  & \multicolumn{1}{c}{$MAMBO$} & \multicolumn{2}{c}{Differences} \\
                                     &  $S_{260GHz}$                 & $S_{250GHz}$               &  $\Delta\alpha$ & $\Delta\delta$ \\    
                                     &    (mJy)                      &   (mJy)                    &      $('')$     &     $('')$    \\
\hline
            7                        &  2.87$\pm$ 0.47               &    3.2$\pm$0.7             &       $0$         & $+6$       \\
            9                        &  3.13$\pm$ 0.45               &    4.3$\pm$0.8             &       $-3$        & $-2$         \\
           10                        &  2.97$\pm$ 0.45               &    6.3$\pm$1.0             &       $-3$        & $-2$         \\
           16                        &  4.11$\pm$ 0.43               &    5.6$\pm$0.7             &       $0$         & $-1$         \\
      {\tiny MM134543+145317}        &  1.3$\pm$ 0.45                &    3.0$\pm$0.7             &       $+3$        & $-2$    \\                           
\hline
\end{tabular}
\label{tab:comp}
\end{table}

\subsection{Nature of the MAMBO sources}\label{nat}

As already mentioned, the field around GJ526 was mapped in 2007 with MAMBO. Five sources were found quasi-aligned and 
within $100''$ from the star GJ526 coordinates, and they were reported in Table~2 of \citet{Lest09}. They were hypothesized to be the clumps 
of a large, broken disc seen almost edge-on surrounding this star. 
With our new NIKA2 observations, four  of these five MAMBO sources
were detected anew in the NIKA2 map at 1.15~mm, and 
are IDs~7, 9, 10, and 16 in Table~\ref{tab:Tab_resul_1mm} herein. The fifth MAMBO source $-$ MM134543+145317, a 4.3$\sigma$ MAMBO detection
in Table~2 of \citet{Lest09} $-$ was also identified  in the  1.15-mm NIKA2 map but only at the level of 2.8$\sigma$ and 
so  was not retained in our final NIKA2 source list.  Such a difference in detection levels would arise from a S/N 
underestimate and overestimate of $0.75\sigma$ by NIKA2 and MAMBO, respectively. 

Astrometric and photometric comparisons between the MAMBO map
and the NIKA2 map, which is 40\% deeper, are given in Table~\ref{tab:comp} herein.  
The coordinate differences between the four MAMBO sources at epoch 2007 (Table~2 of \citealt{Lest09}) 
and their NIKA2 counterparts at epoch 2017  
in  Table~\ref{tab:Tab_resul_1mm} herein  are only a few arcseconds, and so  
 much less than the displacement of $22.6''$ in the south-east direction inferred from the star motion.  
Rather, these coordinates differences are consistent with telescope pointing errors (rms$ < 3''$, see Table~18 of \citealp{Pero20}) 
combined with astrometric errors for sources with low S/Ns.
Thus, we conclude that the MAMBO sources have not moved in concert with the star 
and are not clumps of a disc associated 
with GJ526. They must instead be  SMGs in the background, probably unlensed owing to their low flux density level.

The MAMBO flux densities in Table~2 of \citet{Lest09} 
are comparable with the NIKA2 flux densities at 1.15~mm in Table~\ref{tab:Tab_resul_1mm} herein
to better than 1.5 times the quadratically combined uncertainties of the
two measurements, except for source ID~10 for which the difference is 3.1 times
the combined uncertainty.
This source would have faded from $6.3\pm1$mJy at epoch 2007 to $2.97\pm0.62$mJy ten years later
if this difference were real. This is physically unrealistic for an SMG and we favour calibration difficulties with MAMBO and/or NIKA2, inherent to low S/N sources.
It should be noted that the differences in flux densities due to differences in reference frequencies of the two cameras, 250~GHz for MAMBO 
and  260~GHz for NIKA2, are negligible. 

\subsection{Counterparts in the optical, infrared and radio domains}\label{counterparts}

To search for counterparts to the NIKA2 sources in the optical, mid-infrared, far-infrared, and radio domains, we 
inspected images of the Sloan Digital Sky Survey (SDSS), the Wide-field Infrared Survey Explorer (WISE) \citep{Wrig10}, 
the Herschel Observatory  \citep{Grif10, Potg10}
and the NRAO VLA Sky Survey (NVSS)  \citep{Cond98}. Owing to the NIKA2 position uncertainties, 
we found no convincing counterpart in SDSS, WISE, or NVSS.
In NVSS, with a 3$\sigma$ limit of 1.5~mJy at 1.4~GHz,  this is not  surprising unless it is from  an AGN
 
More interesting were our findings in  the Herschel archive. 
Seven NIKA2 sources have supplemental photometric data from SPIRE (250~$\mu$m, 350~$\mu$m and 500~$\mu$m) and two sources have PACS data 
(100~$\mu$m and 160~$\mu$m). For SPIRE,
the flux densities of three sources (IDs 7, 10, and 18) are taken from the Herschel/SPIRE Point Source Catalogue (HSPSC)  \citep{Schu17}
\footnote{https://irsa.ipac.caltech.edu/Missions/herschel.html} and are reported in Table~\ref{tab:SDSS_WISE_SPIRE} herein,
along with their HSPSC identifiers. We also uploaded the SPIRE observation ID$\_$obs 1342234791 of
the Herschel Open-Time Key project DEBRIS \citep{Matt10} to determine the flux densities of three additional sources
in Table~\ref{tab:SDSS_WISE_SPIRE} (IDs~14, 17, and 21). Their coordinate differences between SPIRE and NIKA2 are, at most, 1.5 times the
quadratically combined astrometric uncertainties of the two instruments ($\sim 5''$). Additionally,
we determined SPIRE upper limits for
the other NIKA2 sources since they are within the SPIRE maps of this programme and not close 
to the map border of lesser quality. We measured a representative rms in an area where there is no apparent source
in the maps to estimate the SPIRE $3\sigma$ upper limits of 34~mJy, 33~mJy, and  31~mJy
at $250\mu$m, $350\mu$m, and $500\mu$m, respectively. 
For PACS, we uploaded the observation ID$\_$obs 1342213079 of the same Open-Time Key project
and found that ID~7 is detected and ID~10 is not. 
We estimated the flux densities of ID~7  using aperture photometry  (radius=30$''$)
and found $S_{\nu}=10\pm2$~mJy at 100~$\mu$m and $S_{\nu}=52\pm3$~mJy at 160~$~\mu$m.
The $3\sigma$ upper limits of ID~10 from this observation are 6~mJy at 100~$\mu$m and 9~mJy at 160~$\mu$m.
The other NIKA2 sources are not part of the PACS maps (ID~18  is at the border
of the maps and the data are unusable).

\begin{table}[]
\centering
\caption{ Herschel/SPIRE photometry }
\begin{tabular}{c|ccc|l}
\hline \hline
\multicolumn{1}{c|}{ID} & \multicolumn{4}{c}{$SPIRE$ }  \\
\hline
            & 250$\mu$m       & 350$\mu$m    & 500$\mu$m    & HSPSC ID    \\
              &  (mJy)          & (mJy)        & (mJy)        &             \\
 \hline
  7           &  65.6$\pm$7.0   & 59.9$\pm$7.4 & 30.9$\pm$6.9 & J1345.65+1451.7 \\
 10       &  32.3$\pm$6.0   & 35.8$\pm$7.8 & 31.6$\pm$7.6 & J1345.68+1454.3 \\
 14           &  49.4$\pm$9.3   & 40.5$\pm$9.1 & 27.4$\pm$6.5 &             \\
 17           &  33.1$\pm$6.5   & 29.8$\pm$5.4 &    $<23$     &             \\
 18       &  77.9$\pm$6.9   & 62.2$\pm$7.2 & 29.1$\pm$7.5 & J1345.85+1451.9   \\
 21           &  45.8$\pm$9.6   & 31.2$\pm$7.2 &    $<23$     &             \\
\hline
\end{tabular}
\label{tab:SDSS_WISE_SPIRE}
\end{table}

\subsection{Background source number count in the GJ526 field}\label{overdensity}

In Sect.\,\ref{nat}, we concluded that there are ten
sources unambiguously detected in the 1.15-mm NIKA2 map with flux densities larger than 2~mJy. 
This can be usefully compared  to the cumulative source number counts in other fields
at a similar wavelength. 
The COSMOS field and the Lockman Hole North field have been surveyed at 1.2~mm  with MAMBO/IRAM 30m  
\citep{Grev04,Bert07,Lind11} and at 1.1~mm  with AzTEC/ASTE  \citep{Scot10,Hats11}. Also, part of the GOODS field 
has been surveyed at 1.1mm with ALMA \citep{Fran18}.

The noise rms in the NIKA2 maps is not uniform, and increases radially from the centre, 
as shown in Fig.\,\ref{fig:map_rms}, because of the scanning strategy used for the observations. 
However, as it can be seen in this figure, the noise rms increases by less than 10\% within 
a radial distance of $150''$ from the map centre and so we first consider this central part of 
the map as the region of  interest with a quasi-uniform noise rms.  
Practically, the statistical uncertainty of the flux density of a source
in this central region is $\sim 0.50$mJy/beam at 1.15~mm. Thus, the detection threshold of $4\sigma$ 
corresponds to a flux density of 2~mJy at this wavelength.
\citet{Lind11} combined their observations with other single dish studies to report the cumulative source number count
 $N(S_{1.2mm} \ge 2.0{\rm mJy})=500\pm200$~sources/deg$^2$.  \citet{Fran18} report  
the lower formal value $N(S_{1.1mm} \ge 1.8{\rm mJy})=209_{-119}^{+178}$~sources/deg$^2$ with ALMA at a similar wavelength. 
It is known that the number counts derived from interferometric data 
are lower than the number counts derived from single-dish data because of source blending \citep{Kari13, Beth17}.
Therefore, consistently with our NIKA2 observations, we adopted the single dish count of \citet{Lind11} for our probability calculation.  
There are  six sources (IDs~7, 9, 10, 13, 16, and 18) that are unambiguously detected (S/N~$>$~4) in
the central region of the 1.15-mm map with quasi-uniform rms ($r < 150''$), while 2.7 sources are expected 
with the density  500~sources/deg$^2$.
The Poisson probability of exactly six sources with $S_{1.2mm} > 2$~mJy being found in this region, given 
this  mean surface density,  is as low as  3.7\%. 
It may be argued that the likelihood of finding an excess of sources in the GJ526 field is actually
enhanced because this field is one in a collection of 50 other similar-sized fields that were mapped in
the initial MAMBO survey \citep{Lest06,Lest09}. Although it is relevant to recall this context, a clear-cut conclusion
cannot be drawn because the other fields were observed with significantly less integration times than GJ526, and so are less deep. 
We discuss now the more intriguing spatial distribution of the sources of this overdensity along a filament-like structure.

\subsection{Filament-like structure} \label{Filam}

\begin{figure}
   \centering
     \resizebox{7cm}{!}{\includegraphics[scale=0.5]{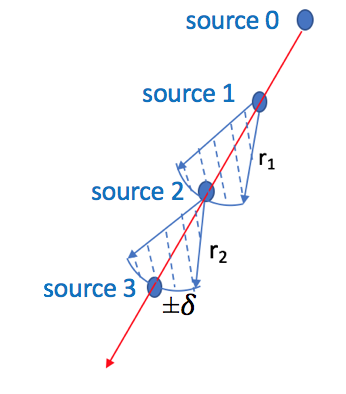} }
   \caption{Estimation of the probability for the chance alignment of sources.}
   \label{fig:sketch}
\end{figure}

Now using the whole map, it is remarkable that seven sources (IDs 3, 6, 9, 10, 16, 18, and 25)
are spatially distributed along an arc in projection on the sky across the whole map, 
as highlighted in Fig.\,\ref{fig:map_1mm}.  
The first six sources are unambiguously detected  with 
flux densities larger than 2~mJy at 1.15~mm, while the last one (ID~25) is 
 a  $3.3\sigma$ possible detection with a flux density of $2.61\pm0.79$~mJy at this wavelength (Table~\ref{tab:Tab_resul_1mm}).  
We can estimate the probability that such a quasi-alignment of sources occurs by chance using small, consecutive sectors  all along the arc with the small opening angle $2 \delta$, as sketched in Fig.\,\ref{fig:sketch},
and in which the averaged source number surface density $\mu$ is $N(S_{1.2mm}\ge2$mJy)~$=500$~sources/degree$^2$, 
as referenced in Sect.\,\ref{overdensity}. 
For  any triplet of sources along the arc, the probability that the third source does not deviate 
more than $\pm \delta$ in degrees from alignment
with the first two sources is the Poisson probability 
$\mu \times \pi r^2 \times 2 \delta/360^{\circ})\exp(\mu \times \pi r^2 \times 2 \delta/360^{\circ}$
that exactly one source is found within the sector centred 
on the second source, and having a radius $r$, which is the distance between the second  and third sources.
In applying this formula for the first triplet in the NIKA2 map (source IDs~3, 6, and 9), the probability of 
a chance occurrence of quasi-alignment within $\pm 10^{\circ}$ is  
$\sim \rm{500~sources/deg}^2 \times \pi (70''/3600'')^2 \times 20^{\circ}/360^{\circ} =3.3\%$ 
with $r_1=70''$ for the distance between source IDs 6 and 9. 
Similarly, the probability is 0.6\% for the second triplet (IDs~6, 9, and 10) with $r_2=30''$ between source IDs 9 and 10,
 it is 10.1\% for the third triplet (IDs 9, 10, and 16) with $r_3=130''$, it is 4.1\% for the fourth triplet (IDs~10, 16, and 18) with
$r_4=80''$, and it is 13.0\% for the fifth triplet (IDs~16, 18, and 25) with $r_5=150''$. 
These triplets defined five consecutive sectors (only two are represented in Fig.\,\ref{fig:sketch}) that are independant, and  
so the resulting probability is the product of the five probabilities estimated above, that is
$\sim \prod\limits_{i=1}^{5} \mu \pi r_i^2 2\delta/360^{\circ}$ (the exponential is approximatively unity in our application), 
and this combined probability 
is  as low as $1.1 \times 10^{-7}$.  Therefore, it is very  unlikely that this filament-like structure has occurred by chance. 
This is true whatever is our concern that 50 similar-sized fields have been observed in the initial MAMBO survey.
In fact, to account for this, the probability can be boosted 50 times and is
 $5.5 \times 10^{-6}$ (i.e. still very low). 

\section{Spectral energy distribution}\label{sed}


Physical quantities of an SMG can be estimated from a model of its SED fit to photometric data.
In an SMG, copious amounts of diffuse small dust grains absorb the intense UV radiation field of massive stars 
and re-radiate in the far-infrared. Its  intrinsic SED peaks at a wavelength corresponding to dust temperature 
typically between 20~K and 60~K, and has a slope steeper 
than the Rayleigh-Jeans regime because of the dust emissivity in the (sub-)millimetre domain. 
 For an SMG with bolometric luminosity $L_{bol}$ at redshift $z$, 
its intrinsic SED $f_{\nu}$ is shifted towards longer wavelengths and the detectable 
flux density at an observed frequency $\nu$  is  \citep[eq. 3]{Blai02} :

\begin{equation}
 S_{\nu}= {{1+z} \over {4 \pi D_L^2}} L_{bol} { {f_{\nu(1+z)}} \over {\int f_{\nu'} d\nu'}} 
\label{eq:sed},
\end{equation}

\noindent where $D_L$ is the luminosity distance to redshift $z$ \citep[][eq. 15.3.24]{Wein72} and can be computed using 
the on-line calculator by \citet{Wri06}
 \footnote{https://ned.ipac.caltech.edu/help/cosmology$_-$calc.html} in standard $\Lambda$CDM cosmology
\footnote{$H_0=67.4$km/s/Mpc and $\Omega_m=0.315$}.
In this equation, the intrinsic SED $f_{\nu}$ is from a modified black-body 
to account for dust emissivity in the far-infrared and sub-millimetre. In other words,{\it } 
it is a Planck function at dust temperature $T_{dust}$ multiplied by
the attenuation factor $1-exp(-(\nu/\nu_c)^{\beta})$ with  the critical frequency $\nu_c$ at which the dust opacity reaches unity.
In our study, we fixed $\nu_c$ to the standard value 3000~GHz (100~$\mu$m) \citep{Blai03}.

 \begin{figure*}
    \centering
      \resizebox{20cm}{!}{\includegraphics[scale=1.0, angle=0]{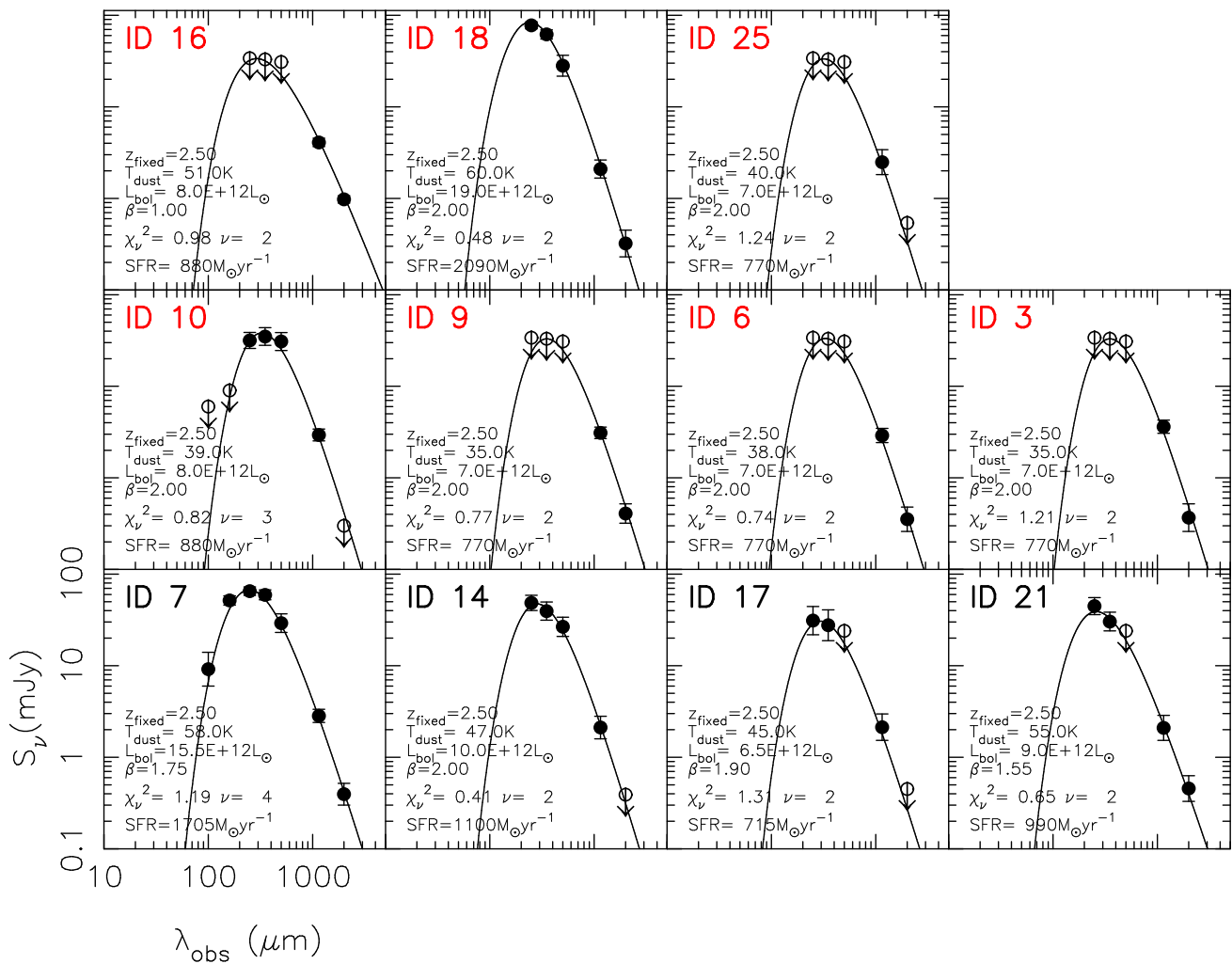}}
      \caption{SEDs of all NIKA2 sources of interest. Each SED is from a single-component modified black-body adjusted 
to the Herschel and NIKA2 data by varying the dust temperature $T_{dust}$, the bolometric luminosity $L_{bol}$, and 
the dust emissivity index $\beta$,
and keeping the redshift fixed at $z=2.5$ for all sources. Empty markers mark the upper limits. 
The seven sources in the candidate cosmic filament discussed in this study are labelled in red. $\chi_{\nu}^2$
 is reduced with the number of degrees of freedom $\nu$ (see details in Sect.\,\ref{sed}).} 
     \label{fig:SED_SPIRE}
 \end{figure*}

For our sources, photometry data were limited to the flux densities from NIKA2 and Herschel  
in Tables\,\ref{tab:Tab_resul_1mm} and \ref{tab:SDSS_WISE_SPIRE}, and only to the SPIRE upper limits for some of them.
Fitting the SED of each source to such a limited amount of data cannot yield a reliable photometric redshift 
because of the known degeneracy between redshift and 
dust temperature when only far-infrared and sub-millimetre  
data are available 
(e.g. Sect.\, 2.10.1 in \citet{Blai02}). Instead, we made the most of the data at our disposal
in testing whether or not they are
compatible with the hypothesis that all sources are at the same redshift. To this end, we constrained
$T_{dust}$ to the range from 20~K to 60~K 
\citep[e.g.][]{Chap05, Stra16}, and constrained all sources to be at the same median redshift $z=2.5$ of the 
phenomenological model of SMGs by \citet{Beth15} for a survey depth 
similar to our NIKA2 observations at 1.15~mm. Figure\,\ref{fig:SED_SPIRE} 
shows that the photometric data are compatible with the
hypothesis that all sources are at the same redshift, and thus 
are in the same cosmic structure. This remains true whether or not flux boosting is applied. Evidently,  
it is only when their spectroscopic redshifts are measured that this hypothesis could be confirmed.   
As far as the three parameters varied in the model, we provide their acceptable ranges for each source in  Fig.\,\ref{fig:SED_SPIRE} ; it is
from 26~K to 60~K for $T_{dust}$ (keeping in mind our initial constraint), it is
from $2.5\times10^{12}$~L$_{\odot}$ to $23\times10^{12}$~L$_{\odot}$ for $L_{bol}$, and  it is from 1.0 to  2.0 for $\beta$.
These luminosities are in the upper range of star-forming ultra-luminous infrared galaxies (ULIRG) in the local universe ($10^{12} \le                                L_{IR}/L_{\odot} \le 10^{13}$~L$_{\odot}$). The brightest source, ID~18, is even formally
in the range of hyper-luminous infrared galaxies (HyLIG).
In addition, in  Fig.\,\ref{fig:SED_SPIRE} we provide the star formation rates (SFRs) between 330 and 2255~M$_{\odot}$/yr, as
 derived from the luminosity L$_{bol}$ using  SFR(M$_{\odot}$/yr)=$1.1 \times 10^{-10} \times$ L$_{bol}$(L$_{\odot}$) \citep{Hayw14}.

\section{Discussion} \label{diss}

The distinctive quasi-alignment of the seven sources, IDs~3, 6, 9, 10, 16, 18, and 25, across the 1.15-mm NIKA2 map 
is qualitatively reminiscent of the imprint of a cosmic filament
in cosmological hydrodynamical simulations 
(e.g.  Illustris \citep{Spri18}, SIMBA \citep{Dave19},  
NewHorizon \citep{Dubo21}, and Uchuu \citep{Ishi21}). 
Quantitatively, these simulations provide spatial gauges that characterize the distribution of dark matter haloes 
 that host galaxies. These gauges can be  usefully compared to our observations. 
First,  it is noticeable that the candidate filament  stretches across the whole NIKA2 maps $560''$ in size 
and so extends at least over $\sim$4~Mpc (comoving) using the scaling factor of 8.3"/kpc, which depends only weakly on
redshift for $1 < z < 6$~\footnote{https://ned.ipac.caltech.edu/help/cosmology$_-$calc.html}. 
Such an observed extent is larger than the size of a typical cluster of galaxies and is comparable to cosmic filaments
in cosmological simulations where they extend from 3~Mpc to 15~Mpc (see 
e.g. NewHorizon snapshot at $z=2$ in Figure~1 of \citet{Dubo21}). Second,  it is  noticeable 
that the five angular separations $r_i$ 
between our NIKA2 sources  given in Sect.\,\ref{Filam} (completing with $r_0=20''$ between IDs~3 and 8) correspond
to linear separations between $\sim$0.25~Mpc and $\sim$1.25~Mpc, using the same  scaling factor. 
Such  observed separations are also comparable to cosmological simulations
 where  haloes hosting galaxies are typically separated by 0.1 to 1.5~Mpc (see e.g. the NewHorizon snapshot mentioned above). 
This match of spatial gauges between our NIKA2 observations and cosmological simulations supports our suggestion 
that the seven SMGs, which are quasi-aligned, bright (L$_{FIR}~>~10^{12}$L$_{\odot}$) ,
and possibly at the same  redshift, may trace a  cosmic filament.    

A more detailed comparison between our observations and cosmological simulations is not an easy task. 
The first simulation that has been specifically motivated to study SMGs is the Gadget-2 simulation by \citet{Dave10} based on a scenario of  gas-rich satellite
 infall rather than intergalactic gas infall along cosmic filaments. However, the spatial distribution of SMGs of this simulation is not presented.
In a more recent study derived from the SIMBA simulation \citep{Dave19}, \citet{Love21}  also focus on SMGs
and in their Fig.~4, they provide a 140 cMpc sized map of the sky projection of    
 all simulated galaxies with SFR~$>$~20~M$_{\odot}$/yr   and  within $0.1 < z < 10$. 
This simulated map is as deep as 0.25~mJy at  850~$\mu$m (corresponding to SFR~$=$~20~M$_{\odot}$/yr)  and is convolved with the SCUBA2 beam.
Their resulting spatial distribution of SMGs is clearly non-uniform 
and structured along filaments. Further comparison will require a simulation zoomed to match the smaller scale of our NIKA2 maps ($\sim$~5cMpc)
and evolved to match our NIKA2 source redshift, when measured. 

\section{Conclusions}

With the NIKA2 dual-band camera at the 30-metre IRAM telescope, we made two deep maps
 of a relatively large field of $\sim$90~arcminutes$^2$ in the direction of the star GJ526, simultaneously 
in its  1.15-mm and 2.0-mm continuum wavebands  
with a $1\sigma$ sensitivity of 0.45~mJy/beam and 0.15~mJy/beam, respectively, 
and with 9.3 hours of on-source observations owing to the high mapping speed of the NIKA2/IRAM 30-metre system. 
In total, ten  background millijansky sources are unambiguously detected with $S/N>4$ and must be SMGs.

In projection on the sky, the  spatial distribution of seven of these SMGs along a filamentary structure that completely crosses 
both maps is remarkable and has only a low probability of chance occurrence ($1.1 \times 10^{-7}$). 
We show that all available photometric data, though limited,  are compatible 
with a model of their SEDs in which they are at the same redshift $z=2.5$. Of course, this must be confirmed with spectroscopic redshift  measurements. At this stage,
we can only speculate that this structure could trace  a cosmic filament of dark matter, as is predicted 
by theory and apparent in cosmological hydrodynamical simulations.
In assuming the scaling factor of 8.3~kpc$/''$ at high redshift with  
standard cosmology for our maps, the extent of this candidate cosmic filament 
is at least 4~Mpc and the separations
between SMGs are a fraction of Mpc comparable to distances between dark matter haloes
 in high-resolution cosmological simulations.
Again, our interpretation is speculative at this stage and 
a full assessment requires measurements of the spectroscopic redshifts of the SMGs in this candidate filament. 
Deep spectroscopic follow-up observations of their molecular (CO) or atomic (CI) lines redshifted into the ALMA or NOEMA bands  
can be envisioned  \citep{Inou16, Hash18, Soli21, Bing21, Birk21}.
An attempt to optically identify them would require both deeper optical observations and
interferometric millimetre-wavelength continuum observations to improve their positions to sub-arcsecond accuracy
Finally, the NIKA2 Cosmology Legacy Survey underway  
at the IRAM 30-metre telescope \citep{Bing21}, designed to map fields more than 30 times larger than 
the  GJ526 field,  will be of great interest to further probe the spatial distribution of SMGs in the millimetre-wavelength domain
and address their relationship with the network of cosmic filaments of dark matter in simulations. 
In particular, establishing this relationship 
would support the cold gas accretion mechanism to fuel galaxy growth.  This endeavour is also pursued with imaging of diffuse 
Ly~$\alpha$ emission along filamentary structures (e.g \citet{Umeh19}).    

\vspace{2.0cm}

\begin{acknowledgements} This work is based on observations carried out under project 
number 133-17 with the IRAM 30m telescope. IRAM is supported by INSU/CNRS (France), 
MPG (Germany) and IGN (Spain).
We would like to thank the IRAM staff for their support during the
campaigns. The NIKA2 dilution cryostat has been designed and built at
the Institut N\'eel. In particular, we acknowledge the crucial
contribution of the Cryogenics Group, and in particular Gregory Garde,
Henri Rodenas, Jean Paul Leggeri, Philippe Camus. This work has been
partially funded by the Foundation Nanoscience Grenoble and the 
LabEx FOCUS ANR-11-LABX-0013. This work is supported by the French
National Research Agency under the contracts ``MKIDS'', ``NIKA'' and 
ANR-15-CE31-0017 and in the framework of the ``Investissements
d’avenir” program (ANR-15-IDEX-02). This work has benefited from 
the support of the European Research Council Advanced Grant ORISTARS 
under the European Union's Seventh Framework Program (Grant
Agreement no. 291294). This research has made use of Herschel data ;
Herschel is an ESA space observatory with science instruments provided by European-led 
Principal Investigator consortia and with important participation from NASA.
This research has made use of the SIMBAD database,
operated at CDS, Strasbourg, France. 
This research has made use of the NASA/IPAC Extragalactic Database (NED),
which is operated by the Jet Propulsion Laboratory, California Institute of Technology,
under contract with the National Aeronautics and Space Administration.
\end{acknowledgements}

%
\bibliographystyle{aa} 
\bibliography{44036corr} 
%



\end{document}